\documentstyle[12pt,fleqn,epsf]{article}
\newcommand{\nn}{\noindent}

\newcommand {\oalf} {\mbox{${\cal O}(\alpha)$}}
\newcommand{\nl}{\nonumber \\}

\newcommand{\bq}{\begin{equation}}
\newcommand{\eq}{\end{equation}}
\newcommand{\ba}{\begin{eqnarray}}
\newcommand{\ea}{\end{eqnarray}}
\newcommand {\oa}{\mbox{${\cal O}(\alpha)$}}
\newcommand{\mr}{\mathrm}
\newcommand{\naive}{na$\ddot{\imath}$ve}
\newcommand{\mathrm}{\rm}
\newcommand{\ww}{\mbox{$\Gamma_W$}}

\newcommand{\AmS}{{\protect\the\textfont2
  A\kern-.1667em\lower.5ex\hbox{M}\kern-.125emS}}
\newcommand{\ltwo}{LEP~2}
\newcommand{\lone}{LEP~1}

\begin{document}
\thispagestyle{empty}
\onecolumn
\date{}
\vspace{-1.4cm}
\begin{flushright}
{CERN--TH.7295/94 \\}
{DESY 94--093 \\}
\end{flushright}
\vspace{.5cm}
\begin{center}
{\Large  \bf
Semi-Analytical Approach to \\
Four-Fermion Production in $e^+ e^-$ Annihilation$\;^{\dag}$}
\vspace*{1.0cm}
\end{center}
\begin{center}
\nn
{\large
D. Bardin$\;^{1,2 }$, $\;$ M. Bilenky$\;^{2,3 }$, $\;$ D. Lehner$\;^3$,
\\
A.~Olchevski$\;^{2,\rm a} \; $ and $\;$ T.~Riemann$\,^3$}
\\
\vspace*{1.0cm}

{\small
\nn
$^1$ 
Theory Division, CERN, CH--1211, Geneva 23, Switzerland
\\
$^2$ 
Joint Institute for Nuclear Research, ul. Joliot Curie 6, \\
RU--141980 Dubna, Moscow Region, Russia
\\
$^3$  
DESY--Institut f\"ur Hochenergiephysik, Platanenallee 6, D--15738
Zeuthen, Germany
}

\vspace*{1.0cm}
\vfill
\end{center}
\vspace*{0.5cm}

\thispagestyle{empty}
\centerline{\bf ABSTRACT}
\vspace*{.3cm}
\nn
A study of the
semi-analytical approach to four-fermion production
in $e^+ e^-$ annihilation is presented.
We classify all possible
four-fermion final states and present results of new calculations
for the `basic' processes with the $WW,ZZ$, and
$ZH$ off-shell production together with some
examples of `background' processes.
The Initial State Radiative
corrections are included for the basic processes.
Several numerical examples are given in the energy range from
\ltwo~up to $\sqrt{s}=1$ TeV.
\vspace*{.5cm}

\bigskip
{\em
\begin{center}
Contribution to the Proceedings of the \\
Zeuthen Workshop on Elementary Particle Theory -- Physics at LEP200
and  Beyond, \\
Teupitz, Germany, 10--15 April, 1994.  \\
\end{center}
}
\vfill
\begin{flushleft}
{
  CERN--TH.7295/94  \\
  DESY 94--093   \\
  June 1994 \\ }
\end{flushleft}
\footnoterule
\nn
{\small
$^{\dag}$ Talk presented by D. Bardin. \\
$^{\rm a}$~Present address: PPE Division, CERN.
{\tt
\begin{tabbing}
email: \= BARDINDY@CERNVM.CERN.CH, bilenky@hades.ifh.de,
\\     \> LEHNER@CONVEX.IFH.DE, OLSHEVSK@VXCERN.CERN.CH,
\\     \> RIEMANN@CERNVM.CERN.CH
\end{tabbing}
} }
%
%
\def\photonatomrightt{\begin{picture}(3,1.5)(0,0)
                               \put(0,-0.75){\tencirc \symbol{2}}
                               \put(1.5,-0.75){\tencirc \symbol{1}}
                               \put(1.5,0.75){\tencirc \symbol{3}}
                               \put(3,0.75){\tencirc \symbol{0}}
                     \end{picture}
                    }
\def\markphotonatomur{\begin{picture}(2,2)(0,0)
                             \put(2,1){\oval(2,2)[tl]}
                             \put(0,1){\oval(2,2)[br]}
                     \end{picture}
                    }
\def\markphotonatomdr{\begin{picture}(2,2)(0,0)
                             \put(1,0){\oval(2,2)[bl]}
                             \put(1,-2){\oval(2,2)[tr]}
                     \end{picture}
                    }

\def\photonatomright{\begin{picture}(3,1.5)(0,0)
                                \put(0,-0.75){\tencircw \symbol{2}}
                                \put(1.5,-0.75){\tencircw \symbol{1}}
                                \put(1.5,0.75){\tencircw \symbol{3}}
                                \put(3,0.75){\tencircw \symbol{0}}
                      \end{picture}
                     }
\def\photonatomupt{\begin{picture}(1.5,3)(0,0)
                            \put(-0.75,3){\tencirc \symbol{3}}
                            \put(-0.75,1.5){\tencirc \symbol{2}}
                            \put(0.75,1.5){\tencirc \symbol{0}}
                            \put(0.75,0){\tencirc \symbol{1}}
                  \end{picture}
                 }
\def\photonatomup{\begin{picture}(1.5,3)(0,0)
                             \put(-0.75,3){\tencircw \symbol{3}}
                             \put(-0.75,1.5){\tencircw \symbol{2}}
                             \put(0.75,1.5){\tencircw \symbol{0}}
                             \put(0.75,0){\tencircw \symbol{1}}
                   \end{picture}
                  }
\def\gluonatomrightt{\begin{picture}(3.5,3.25)(0,0)
                               \put(2.5,0){\tencirc \symbol{9}}
                               \put(1,0){\tencirc \symbol{10}}
                               \put(1,3){\tencirc \symbol{3}}
                               \put(2.5,3){\tencirc \symbol{0}}
                    \end{picture}
                   }
\def\gluonatomright{\begin{picture}(3.5,3.25)(0,0)
                                \put(2.5,0){\tencircw \symbol{9}}
                                \put(1,0){\tencircw \symbol{10}}
                                \put(1,3){\tencircw \symbol{3}}
                                \put(2.5,3){\tencircw \symbol{0}}
                     \end{picture}
                    }
\def\gluonatomupt{\begin{picture}(3.25,3.5)(0,0)
                            \put(0,1){\tencirc \symbol{10}}
                            \put(0,2.5){\tencirc \symbol{11}}
                            \put(3,2.5){\tencirc \symbol{0}}
                            \put(3,1){\tencirc \symbol{1}}
                 \end{picture}
                }
\def\gluonatomup{\begin{picture}(3.25,3.5)(0,0)
                             \put(0,1){\tencircw \symbol{10}}
                             \put(0,2.5){\tencircw \symbol{11}}
                             \put(3,2.5){\tencircw \symbol{0}}
                             \put(3,1){\tencircw \symbol{1}}
                  \end{picture}
                 }
\def\photonrightt{\begin{picture}(30,1.5)(0,0)
                    \multiput(0,0)(3,0){10}{\photonatomrightt}
                 \end{picture}
                }
\def\photonrightthalf{\begin{picture}(15,1.5)(0,0)
                    \multiput(0,0)(3,0){5}{\photonatomrightt}
                 \end{picture}
                }
\def\photonright{\begin{picture}(30,1.5)(0,0)
                     \multiput(0,0)(3,0){10}{\photonatomright}
                  \end{picture}
                 }
\def\markphotonur{\begin{picture}(30,30)(0,0)
                     \multiput(0,0)(2,2){15}{\markphotonatomur}
                  \end{picture}
                 }
\def\markphotondr{\begin{picture}(30,30)(0,0)
                     \multiput(0,0)(2,-2){15}{\markphotonatomdr}
                  \end{picture}
                 }
\def\photonrighthalf{\begin{picture}(30,1.5)(0,0)
                     \multiput(0,0)(3,0){5}{\photonatomright}
                  \end{picture}
                 }
\def\photonupt{\begin{picture}(1.5,30)(0,0)
                 \multiput(0,0)(0,3){10}{\photonatomupt}
              \end{picture}
             }
\def\photonupthalf{\begin{picture}(1.5,15)(0,0)
                     \multiput(0,0)(0,3){5}{\photonatomupt}
                  \end{picture}
                 }
\def\photonup{\begin{picture}(1.5,30)(0,0)
                  \multiput(0,0)(0,3){10}{\photonatomup}
               \end{picture}
              }
\def\photonuphalf{\begin{picture}(1.5,15)(0,0)
                      \multiput(0,0)(0,3){5}{\photonatomup}
                   \end{picture}
                  }
\def\gluonrightt{\begin{picture}(30,3.25)(0,0)
                    \multiput(0,0)(3,0){10}{\gluonatomrightt}
                 \end{picture}
                }
\def\gluonrightthalf{\begin{picture}(15,3.25)(0,0)
                        \multiput(0,0)(3,0){5}{\gluonatomrightt}
                     \end{picture}
                    }
\def\gluonrightthhalf{\begin{picture}(7.5,3.25)(0,0)
                         \multiput(0,0)(3,0){3}{\gluonatomrightt}
                      \end{picture}
                     }
\def\gluonright{\begin{picture}(30,3.25)(0,0)
                     \multiput(0,0)(3,0){10}{\gluonatomright}
                  \end{picture}
                 }
\def\gluonrighthalf{\begin{picture}(15,3.25)(0,0)
                         \multiput(0,0)(3,0){5}{\gluonatomright}
                      \end{picture}
                     }
\def\gluonrighthhalf{\begin{picture}(7.5,3.25)(0,0)
                         \multiput(0,0)(3,0){3}{\gluonatomright}
                     \end{picture}
                     }
\def\gluonupt{\begin{picture}(3.25,30)(0,0)
                 \multiput(0,0)(0,3){10}{\gluonatomupt}
              \end{picture}
             }
\def\gluonupthalf{\begin{picture}(3.25,15)(0,0)
                     \multiput(0,0)(0,3){5}{\gluonatomupt}
                  \end{picture}
                 }
\def\gluonup{\begin{picture}(3.25,30)(0,0)
                 \multiput(0,0)(0,3){10}{\gluonatomup}
              \end{picture}
             }
\def\gluonuphalf{\begin{picture}(3.25,15)(0,0)
                      \multiput(0,0)(0,3){5}{\gluonatomup}
                   \end{picture}
                  }
\def\fermionup{\begin{picture}(1,30)(0,0)
                     \put(0,0){\vector(0,1){15}}
                     \put(0,15){\line(0,1){15}}
               \end{picture}
              }
\def\fermionuphalf{\begin{picture}(1,15)(0,0)
                         \put(0,0){\vector(0,1){7.5}}
                         \put(0,7.5){\line(0,1){7.5}}
                   \end{picture}
                  }
\def\fermionupp{\begin{picture}(1,30)(0,0)
                     \put(0,0){\vector(1,2){7.5}}
                     \put(7.5,15){\line(1,2){7.5}}
               \end{picture}
              }
\def\fermionupphalf{\begin{picture}(1,15)(0,0)
                         \put(0,0){\vector(1,2){3.75}}
                         \put(3.75,7.5){\line(1,2){3.75}}
                   \end{picture}
                  }
\def\fermiondown{\begin{picture}(1,30)(0,-30)
                       \put(0,0){\vector(0,-1){15}}
                       \put(0,-15){\line(0,-1){15}}
                 \end{picture}
                }
\def\fermiondownhalf{\begin{picture}(1,15)(0,-15)
                           \put(0,0){\vector(0,-1){7.5}}
                           \put(0,-7.5){\line(0,-1){7.5}}
                     \end{picture}
                    }
\def\fermionleft{\begin{picture}(30,1)(0,0)
                       \put(30,0){\vector(-1,0){15}}
                       \put(15,0){\line(-1,0){15}}
                 \end{picture}
                }
\def\fermionlefthalf{\begin{picture}(15,1)(0,0)
                           \put(15,0){\vector(-1,0){7.5}}
                           \put(7.5,0){\line(-1,0){7.5}}
                     \end{picture}
                    }
\def\fermionright{\begin{picture}(30,1)(0,0)
                        \put(0,0){\vector(1,0){15}}
                        \put(15,0){\line(1,0){15}}
                  \end{picture}
                 }
\def\fermionrighthalf{\begin{picture}(15,1)(0,0)
                            \put(0,0){\vector(1,0){7.5}}
                            \put(7.5,0){\line(1,0){7.5}}
                      \end{picture}
                     }
\def\fermionul{\begin{picture}(15,15)(0,0)
                        \put(0,0){\vector(-1,1){7.5}}
                        \put(-7.5,7.5){\line(-1,1){7.5}}
                  \end{picture}
                 }
\def\fermionulhalf{\begin{picture}(7.5,7.5)(0,0)
                        \put(0,0){\vector(-1,1){3.75}}
                        \put(-3.75,3.75){\line(-1,1){3.75}}
                  \end{picture}
                 }
\def\fermionur{\begin{picture}(15,15)(0,0)
                        \put(-15,-15){\vector(1,1){7.5}}
                        \put(-7.5,-7.5){\line(1,1){7.5}}
                  \end{picture}
                 }
\def\fermionurhalf{\begin{picture}(7.5,7.5)(0,0)
                        \put(-7.5,-7.5){\vector(1,1){3.75}}
                        \put(-3.75,-3.75){\line(1,1){3.75}}
                  \end{picture}
                 }
\def\fermiondl{\begin{picture}(15,15)(0,0)
                        \put(15,15){\vector(-1,-1){7.5}}
                        \put(7.5,7.5){\line(-1,-1){7.5}}
                  \end{picture}
                 }
\def\fermiondlhalf{\begin{picture}(7.5,7.5)(0,0)
                        \put(7.5,7.5){\vector(-1,-1){3.75}}
                        \put(3.75,3.75){\line(-1,-1){3.75}}
                  \end{picture}
                 }
\def\fermiondr{\begin{picture}(15,15)(0,0)
                        \put(0,0){\vector(1,-1){7.5}}
                        \put(7.5,-7.5){\line(1,-1){7.5}}
                  \end{picture}
                 }
\def\fermiondrhalf{\begin{picture}(7.5,7.5)(0,0)
                        \put(0,0){\vector(1,-1){3.75}}
                        \put(3.75,-3.75){\line(1,-1){3.75}}
                  \end{picture}
                 }
\def\fermionull{\begin{picture}(30,15)(0,0)
                        \put(0,0){\vector(-2,1){15}}
                        \put(-15,7.5){\line(-2,1){15}}
                  \end{picture}
                 }
\def\fermionullhalf{\begin{picture}(15,7.5)(0,0)
                        \put(0,0){\vector(-2,1){7.5}}
                        \put(-7.5,3.75){\line(-2,1){7.5}}
                  \end{picture}
                 }
\def\fermionurr{\begin{picture}(30,15)(0,0)
                        \put(-30,-15){\vector(2,1){15}}
                        \put(-15,-7.5){\line(2,1){15}}
                  \end{picture}
                 }
\def\fermionurrhalf{\begin{picture}(15,7.5)(0,0)
                        \put(-15,-7.5){\vector(2,1){7.5}}
                        \put(-7.5,-3.75){\line(2,1){7.5}}
                  \end{picture}
                 }
\def\fermiondrr{\begin{picture}(30,15)(0,0)
                        \put(0,0){\vector(2,-1){15}}
                        \put(15,-7.5){\line(2,-1){15}}
                  \end{picture}
                 }
\def\fermiondrrhalf{\begin{picture}(15,7.5)(0,0)
                        \put(0,0){\vector(2,-1){7.5}}
                        \put(7.5,-3.75){\line(2,-1){7.5}}
                  \end{picture}
                 }
\def\fermiondll{\begin{picture}(30,15)(0,0)
                        \put(30,15){\vector(-2,-1){15}}
                        \put(15,7.5){\line(-2,-1){15}}
                  \end{picture}
                 }
\def\fermiondllhalf{\begin{picture}(15,7.5)(0,0)
                        \put(15,7.5){\vector(-2,-1){7.5}}
                        \put(7.5,3.75){\line(-2,-1){7.5}}
                  \end{picture}
                 }
\def\gaugebosonright{\begin{picture}(30,1)(0,0)
                            \put(0,0){\line(1,0){0.75}}
                            \multiput(2.25,0)(3,0){9}{\line(1,0){1.5}}
                            \put(29.25,0){\line(1,0){0.75}}
                     \end{picture}
                    }
\def\gaugebosonrighthalf{\begin{picture}(15,1)(0,0)
                            \put(0,0){\line(1,0){0.75}}
                            \multiput(2.25,0)(3,0){4}{\line(1,0){1.5}}
                            \put(14.25,0){\line(1,0){0.75}}
                         \end{picture}
                        }
\def\gaugebosonup{\begin{picture}(1,30)(0,0)
                    \put(0,0){\line(0,1){0.75}}
                    \multiput(0,2.25)(0,3){9}{\line(0,1){1.5}}
                    \put(0,29.25){\line(0,1){0.75}}
                  \end{picture}
                 }
\def\gaugebosonuphalf{\begin{picture}(1,15)(0,0)
                            \put(0,0){\line(0,1){0.75}}
                            \multiput(0,2.25)(0,3){4}{\line(0,1){1.5}}
                            \put(0,14.25){\line(0,1){0.75}}
                      \end{picture}
                     }
\def\gaugebosonur{\begin{picture}(30,15)(0,0)
                            \put(0,0){\line(1,1){8.0}}
                            \put(11,5.5){\line(1,1){8.0}}
                            \put(22,11){\line(1,1){8.0}}
                  \end{picture}
                 }
\def\gaugebosonurhalf{\begin{picture}(15,15)(0,0)
                            \put(0,0){\line(1,1){15.0}}
                  \end{picture}
                 }
\def\gaugebosonul{\begin{picture}(30,15)(0,0)
                            \put(0,0){\line(-1,1){8.0}}
                            \put(11,5.5){\line(-1,1){8.0}}
                            \put(22,11){\line(-1,1){8.0}}
                  \end{picture}
                 }
\def\gaugebosonulhalf{\begin{picture}(15,15)(0,0)
                            \put(0,0){\line(-1,1){15.0}}
                  \end{picture}
                 }
\def\gaugebosondr{\begin{picture}(30,15)(0,0)
                            \put(0,0){\line(2,-1){8.0}}
                            \put(11,-5.5){\line(2,-1){8.0}}
                            \put(22,-11){\line(2,-1){8.0}}
                  \end{picture}
                 }
\def\gaugebosondrhalf{\begin{picture}(15,15)(0,0)
                            \put(0,0){\line(1,-1){15}}
                  \end{picture}
                 }
\def\gaugebosondl{\begin{picture}(30,15)(0,0)
                            \put(0,0){\line(-2,-1){8.0}}
                            \put(11,-5.5){\line(-2,-1){8.0}}
                            \put(22,-11){\line(-2,-1){8.0}}
                  \end{picture}
                 }
\def\gaugebosondlhalf{\begin{picture}(15,15)(0,0)
                            \put(0,0){\line(-1,-1){15}}
                  \end{picture}
                 }
\def\gaugebosonurr{\begin{picture}(30,15)(0,0)
                            \put(0,0){\line(2,1){8.0}}
                            \put(11,5.5){\line(2,1){8.0}}
                            \put(22,11){\line(2,1){8.0}}
                  \end{picture}
                 }
\def\gaugebosonurrhalf{\begin{picture}(15,7.5)(0,0)
                            \put(0,0){\line(2,1){15.0}}
                  \end{picture}
                 }
\def\gaugebosondrr{\begin{picture}(30,15)(0,0)
                            \put(0,0){\line(2,-1){8.0}}
                            \put(11,-5.5){\line(2,-1){8.0}}
                            \put(22,-11){\line(2,-1){8.0}}
                  \end{picture}
                 }
\def\gaugebosondrrhalf{\begin{picture}(15,7.5)(0,0)
                            \put(0,0){\line(2,-1){15}}
                  \end{picture}
                 }
\def\gluonatomur{\begin{picture}(4.0,3.25)(0,0)
                       \put(2.4,0.1){\tencircw \symbol{9}}
                       \put(1,0.5){\tencircw \symbol{10}}
                       \put(1,3.5){\tencircw \symbol{3}}
                       \put(2.4,3.1){\tencircw \symbol{0}}
                 \end{picture}
                }
\def\gluonatomdr{\begin{picture}(4.0,3.25)(0,0)
                       \put(2.5,-0.1){\tencircw \symbol{8}}
                       \put(1,-0.5){\tencircw \symbol{11}}
                       \put(1,-3.5){\tencircw \symbol{2}}
                       \put(2.5,-3.1){\tencircw \symbol{1}}
                 \end{picture}
                }
\def\gluonatomurr{\begin{picture}(4.0,3.25)(0,0)
                        \put(2.4,0.1){\tencirc \symbol{9}}
                        \put(1,0.5){\tencirc \symbol{10}}
                        \put(1,3.5){\tencirc \symbol{3}}
                        \put(2.4,3.1){\tencirc \symbol{0}}
                  \end{picture}
                 }
\def\gluonatomdrr{\begin{picture}(4.0,3.25)(0,0)
                        \put(2.5,-0.1){\tencirc \symbol{9}}
                        \put(1,-0.5){\tencirc \symbol{10}}
                        \put(1,-3.5){\tencirc \symbol{3}}
                        \put(2.5,-3.1){\tencirc \symbol{0}}
                  \end{picture}
                 }
\def\gluonur{\begin{picture}(30,3.25)(0,0)
                   \multiput(0,0)(2.7,0.65){11}{\gluonatomur}
             \end{picture}
            }
\def\gluonurhalf{\begin{picture}(15,3.25)(0,0)
                       \multiput(0,0)(2.7,0.65){6}{\gluonatomur}
                 \end{picture}
                }
\def\gluonurhhalf{\begin{picture}(7.5,3.25)(0,0)
                        \multiput(0,0)(2.7,0.65){3}{\gluonatomur}
                  \end{picture}
                 }
\def\gluondr{\begin{picture}(30,3.25)(0,0)
                   \multiput(0,0)(2.7,-0.65){11}{\gluonatomdr}
             \end{picture}
            }
\def\gluondrhalf{\begin{picture}(15,3.25)(0,0)
                       \multiput(0,0)(2.7,-0.65){6}{\gluonatomdr}
                 \end{picture}
                }
\def\gluondrhhalf{\begin{picture}(7.5,3.25)(0,0)
                        \multiput(0,0)(2.7,-0.65){3}{\gluonatomdr}
                  \end{picture}
                 }
\def\gluonurr{\begin{picture}(30,3.25)(0,0)
                    \multiput(0,0)(2.7,0.65){11}{\gluonatomurr}
              \end{picture}
             }
\def\gluonurrhalf{\begin{picture}(15,3.25)(0,0)
                        \multiput(0,0)(2.7,0.65){6}{\gluonatomurr}
                  \end{picture}
                 }
\def\gluonurrhhalf{\begin{picture}(7.5,3.25)(0,0)
                         \multiput(0,0)(2.7,0.65){3}{\gluonatomurr}
                   \end{picture}
                  }
\def\gluondrr{\begin{picture}(30,3.25)(0,0)
                    \multiput(0,0)(2.7,-0.65){11}{\gluonatomdrr}
              \end{picture}
             }
\def\gluondrrhalf{\begin{picture}(15,3.25)(0,0)
                        \multiput(0,0)(2.7,-0.65){6}{\gluonatomdrr}
                  \end{picture}
                 }
\def\gluondrrhhalf{\begin{picture}(7.5,3.25)(0,0)
                         \multiput(0,0)(2.7,-0.65){3}{\gluonatomdrr}
                   \end{picture}
                  }
\def\fermionndl{\begin{picture}(15,15)(0,0)
                      \put(15,15){\line(-1,-1){15}}
                \end{picture}
               }
\def\fermionndlhalf{\begin{picture}(7.5,7.5)(0,0)
                          \put(7.5,7.5){\line(-1,-1){7.5}}
                    \end{picture}
                   }
\def\fermionndr{\begin{picture}(15,15)(0,0)
                      \put(0.0,0.0){\line(1,-1){15}}
                \end{picture}
               }
\def\fermionndrhalf{\begin{picture}(7.5,7.5)(0,0)
                          \put(0.0,0.0){\line(1,-1){7.5}}
                    \end{picture}
                   }

\newenvironment{Feynman}[3]{\begin{center}
                            \setlength{\unitlength}{#3 mm}
                            \begin{picture}(#1)(#2)
                            \thicklines
                           }{\end{picture} \end{center}}

\hyphenation{author another created financial paper re-commend-ed}
\hyphenation{rederi-ved}
\vfil\eject
\section{INTRODUCTION}
\setcounter{page}{1}
While at \lone\ the basic process is two-fermion production
via a single $Z$-boson (or photon) exchange, at \ltwo\ the
typical process will be the `double-resonance' production of four-fermion
final states.
\ltwo\ will operate just in the threshold region
of $W$- or $Z$-pair production and,
if the Higgs-boson mass fulfils $60 \le  M_H \le 100$ GeV,
of $ZH$ production.
Already in the tree level approximation,
the double-resonance physics is much richer and much more interesting
and complex than the single $Z$-boson production.
\par
A specific four-fermion final state can be produced by many
Feynman diagrams with many possible virtual states,
including all the carriers of fermion interactions in the Standard Model:
$\gamma, Z, W^{\pm}$, $g$, and $H$.
We will distinguish between `basic'
diagrams, which contain two potentially resonating virtual states
($W$, $Z$, $H$) in the $s$-channel and
`background' diagrams, which are just the rest.
The general topology of the basic diagrams is shown in figure~1.
The contribution of background diagrams to a given final state is
usually suppressed.
Some background diagrams for $W$-pair production are given in figure~2.
\par
The analytical result for the Born
 on-shell $W$-pair production has been known
for long in the literature~\cite{ONWW}\footnote{
Some approximate
results on the analytical treatment of off-shell $W$-pair production
near threshold can be found in~\cite{GKP83}.}.
The off-shell case was treated in~\cite{MNW86}.
A calculation of off-shell $Z$-pair production was done in~\cite{ONZZ}.
\par
In this paper we present semi-analytical
results for off-shell production
of bosonic pairs ($WW,~ZZ,~ZH$)
including universal lowest-order
Initial State Radiative (ISR) corrections
with soft-photon exponentiation.
For the case of $W$-pairs,
we also present results with complete \oalf\ ISR corrections
and some examples of four-fermion background processes.
\par
The paper is organized as follows:
In the next section, we classify the four-fermion production
processes in $e^+e^-$ annihilation. In section~3
we introduce the notations and present the formulae
for the basic processes.
In section~4, the contributions of the background diagrams
to the simplest final state are characterized.
The ISR corrections are described
in section~5.
In section 6, we present concluding remarks and prospects.
%
\section{A CLASSIFICATION OF FOUR-FERMION\break PROCESSES}
In this section we will classify
the four-fermion production in the Standard Model
\footnote{The classification is done with the use of
CompHEP~\cite{comphep}.}.
The number of Feynman diagrams depends crucially on the
final state.
In general all possible final states can be subdivided into
two classes.
\begin{figure*}
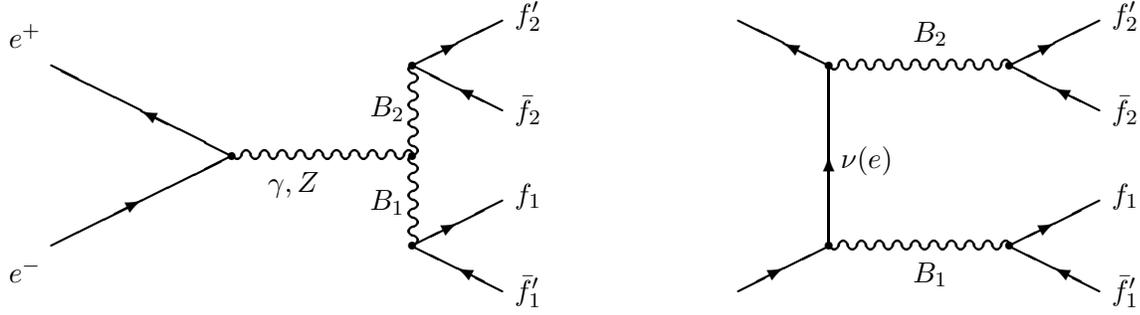

\begin{minipage}[tbh]{7.8cm}{
\begin{center}
\begin{Feynman}{75,60}{0,0}{0.8}
%
\put(30,30){\fermionurr}
\put(30,30){\fermionull}
\put(30,30){\photonright}
\put(30,30){\circle*{1.5}}
\put(60,30){\circle*{1.5}}
\put(60,15){\circle*{1.5}}
\put(60,45){\circle*{1.5}}
\put(75,7.5){\fermionullhalf}
\put(75,22.5){\fermionurrhalf}
\put(60,15){\photonuphalf}
\put(60,30){\photonuphalf}
\put(75,37.5){\fermionullhalf}
\put(75,52.5){\fermionurrhalf}
\small
\put(-07,48){$e^+$}  
\put(-07,09){$e^-$}
\put(36,24){$\gamma, Z$}
\put(53,36.5){$B_2$}  
\put(53,21.0){$B_1$}
\put(77,06){${\bar f}_1^{\prime}$} 
\put(77,22){$ f_1$}
\put(77,36){${\bar f}_2$}
\put(77,52){$ f_2^{\prime}$}
\normalsize
\end{Feynman}
\end{center}
}\end{minipage}
\begin{minipage}[tbh]{7.8cm} {
\begin{center}
\begin{Feynman}{75,60}{0,0}{0.8}
%
\put(30,15){\fermionurrhalf}
\put(30,45){\fermionullhalf}
\put(30,15){\fermionup}
\put(30,45){\photonright}
\put(30,15){\photonright}
\put(30,45){\circle*{1.5}}
\put(30,15){\circle*{1.5}}
\put(60,15){\circle*{1.5}}
\put(60,45){\circle*{1.5}}
\put(75,7.5){\fermionullhalf}
\put(75,22.5){\fermionurrhalf}
\put(75,37.5){\fermionullhalf}
\put(75,52.5){\fermionurrhalf}
\small
\put(32,28){$\nu(e)$}
\put(44,49){$B_2$}  
\put(44,09){$B_1$}
\put(77,06){${\bar f}_1^{\prime}$} 
\put(77,22){$ f_1$}
\put(77,36){${\bar f}_2$}
\put(77,52){$ f_2^{\prime}$}
\normalsize
\end{Feynman}
\end{center}
}\end{minipage}
\caption
{
The basic contributions to off-shell boson-pair production:
{\tt crayfish} and {\tt crab}; $B=W^{\pm},Z,H$.
}
\end{figure*}
\par
The first class comprises
production of up (anti-up) and
anti-down (down) fermion pairs,
\ba
  (U_i~{\bar D_i })~ +~(D_j~{\bar U_j })~,
  \nonumber
\ea
where $i,j$ are generation indices.
The final states produced via virtual $W$-pairs
belong to this class.
We will call these `CC' type final states.
The second class is the production of two fermion-antifermion pairs,
\ba
  (f_i~~{\bar f_i })~ +~(f_j~~{\bar f_j })~,f=U,~D.
  \nonumber
\ea
As it is produced via a pair of two virtual neutral vector
bosons we will call this a final state
of the `NC' type. Obviously these two classes overlap for certain
final states.
\par
The number of Feynman diagrams in the `CC' class is
shown in table~1.
Three different cases occur in the table:

\begin{table}[hbt]
\label{tab1}
\begin{center}
\begin{tabular}{|c|c|c|c|c|c|}
\hline
             &
\raisebox{0.pt}[2.5ex][0.0ex]{${\bar d} u$}
& ${\bar s} c$ & ${\bar e} \nu_{e}$ &
              ${\bar \mu} \nu_{\mu}$ & ${\bar \tau} \nu_{\tau}$   \\
\hline
$d {\bar u}$            &{\it  43}& {\bf 11} &  20 & {\bf 10} & {\bf 10} \\
\hline
$e {\bar \nu}_{e}$      &  20 &  20 &{\it 56}&  18 &  18 \\
\hline
 $\mu {\bar \nu}_{\mu}$ & {\bf 10} & {\bf 10} &  18 & {\it 19} & {\bf 9}  \\
\hline
\end{tabular}
\vspace{.2cm} \\
\caption[]
{
Number of Feynman diagrams contributing to the production of `CC' type
final states.
}
\end{center}
\end{table}

\begin{itemize}
\item[(i)]
    The two produced fermion pairs are different $(i\neq j)$
    and the final state does not contain an $e^{\pm}$
    (numbers in {\bf boldface}). For this case, the number of
    diagrams varies between 9 and 11, depending on the final state's
    neutrino content. The background diagrams for this
    simplest case are shown in figure~2.

\item[(ii)]
    The four-fermion final state contains one $e {\bar \nu}_{e}$-
    or ${\bar e} \nu_{e}$-pair
    (roman numbers); the number of diagrams grows to 18, 19 or 20,
    due to the additional $t$-channel exchange diagrams.

\item[(iii)]
    Two mutually charge-conjugated fermion pairs ($i = j$)
    are produced ({\it italic} numbers). Here, the diagrams may
    contain neutral-boson
    ($Z$, $\gamma$, $H$, gluon) exchanges. One should emphasize that
    this overlaps the `NC' classification.
\end{itemize}

For the final states corresponding to the `NC' class
the number of Feynman diagrams is presented in table~2:

\begin{itemize}
\item[(i)]
The simplest case (numbers in
{\bf boldface}) does not contain electrons or identical
fermions\footnote{
We exclude the Higgs-boson exchange diagrams from the classification
in the tables.}.
\item[(ii)]
With identical fermions $f$ ($f \neq e,\nu_e$),
the number of diagrams (in {\tt typewriter})
grows drastically, since it is
necessary to satisfy the Pauli principle
(i.e. to antisymmetrize the amplitude).
\item[(iii)]
The numbers in romans
correspond to the final states
that include $f=e,\nu_e$ except those covered by item {(iv)}.
The large number of diagrams here is due
to additional $t$-channel diagrams.
\item[(iv)]
The numbers in {\it italics} correspond to final states that are also
present in table~1, case~(iii).
The basic diagrams proceed via both $WW$- and $ZZ$-exchanges.
\end{itemize}

\begin{figure*}
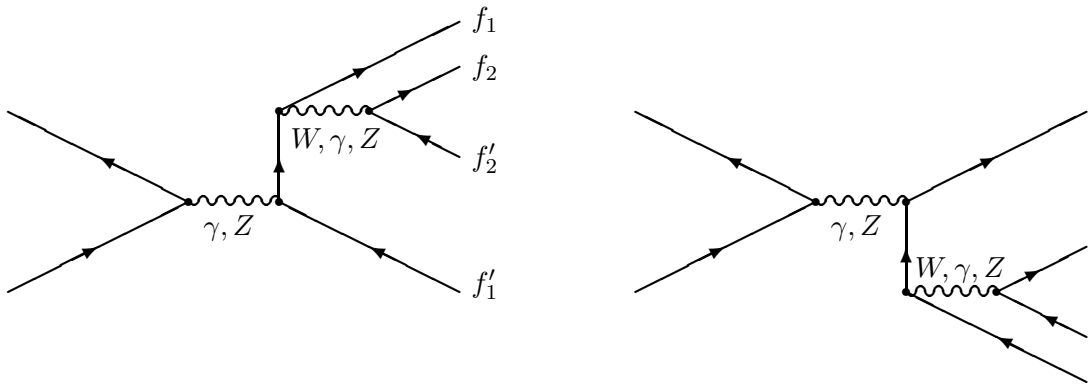

\begin{minipage}[tbh]{7.8cm}{
\begin{center}
\begin{Feynman}{75,60}{-5.0,0}{0.8}
%
\put(20,30){\fermionurr}
\put(20,30){\fermionull}
\put(20,30){\photonrighthalf}
\put(65,15){\fermionull}
\put(20,30){\circle*{1.5}}
\put(35,30){\circle*{1.5}}
\put(35,45){\circle*{1.5}}
\put(50,45){\circle*{1.5}}
\put(35,30){\fermionuphalf}
\put(65,60){\fermionurr}
\put(35,45){\photonrighthalf}
\put(65,52.5){\fermionurrhalf}
\put(65,37.5){\fermionullhalf}
\small
\put(22.5,24.5){$\gamma, Z$}
 \put(37,39){$W,\gamma,Z$}
\put(67,59){$f_1$}
\put(67,51){$f_2$}
\put(67,36){${f}_2^{\prime}$}  
\put(67,15){${f}_1^{\prime}$}  
\normalsize
\end{Feynman}
\end{center}
}\end{minipage}
\begin{minipage}[tbh]{7.8cm}{
\begin{center}
\begin{Feynman}{75,60}{0,0}{0.8}
%
\put(30,30){\fermionurr}
\put(30,30){\fermionull}
\put(30,30){\photonrighthalf}
\put(30,30){\circle*{1.5}}
\put(45,30){\circle*{1.5}}
\put(45,15){\circle*{1.5}}
\put(60,15){\circle*{1.5}}
\put(45,15){\fermionuphalf}
\put(75,00){\fermionull}
\put(75,45){\fermionurr}
\put(45,15){\photonrighthalf}
\put(75,22.5){\fermionurrhalf}
\put(75,07.5){\fermionullhalf}
\small
\put(32.5,24.5){$\gamma, Z$}
 \put(46.5,17.5){$W,\gamma,Z$}
\normalsize
\end{Feynman}
\end{center}
}\end{minipage}
\caption{
Background contributions to off-shell $W$-pair production:
up and down {\tt reindeers}.
}
\end{figure*}
So far we have investigated semi-analytically
only the two simplest cases of
the `CC' and the `NC' classifications.
In section~4 we will present
the production of two different fermion pairs:
\ba
e^+ e^- \rightarrow  (W^+ W^-, W^{\pm}{\bar f}f^{\prime})
\rightarrow
{\bar f}_1^{\prime} f_1 f_2^{\prime} {\bar f}_2,
\nl
f_i,f_j^{\prime} \neq e, \nu_e~.
\nonumber
\label{eqabs}
\ea
\begin{table}[t]
\vspace{-.25cm}
\label{tab2}
\begin{center}
 \begin{tabular}{|c|c|c|c|c|c|c|}
\hline
&
\raisebox{0.pt}[2.5ex][0.0ex]{${\bar d} d$}
&${\bar u} u$
&${\bar e} e$
&${\bar \mu} \mu$
&${\bar \nu}_{e} \nu_{e}$
&${\bar \nu}_{\mu} \nu_{\mu}$
\\
\hline
\raisebox{0.pt}[2.5ex][0.0ex]{${\bar d} d$}
 & {\tt 4$\cdot $16} & {\it 43} & {48}
             & {\bf 24} & 21 & {\bf 10} \\
\hline
${\bar s} s$ & {\bf 32} & {\it 43} & {48}
             & {\bf 24} & {21} & {\bf 10} \\
\hline
${\bar u} u$ & {\it 43} & {\tt 4$\cdot$16} & {48}
             & {\bf 24} & {21} & {\bf 10} \\
\hline
${\bar e} e$ &{48} &{48} & {\tt 4$\cdot$36} &{48}
& {\it 56} & {20}
\\
\hline
${\bar \mu} \mu$  & {\bf 24} & {\bf 24} & {48} & {\tt 4$\cdot$12}
                  & {19} & {\it 19}         \\
\hline
${\bar \tau} \tau$& {\bf 24} & {\bf 24} & {48} & {\bf 24}
                  & {19} & {\bf 10}         \\
\hline
${\bar \nu}_e \nu_{e}$  & {21} & {21} & {\it 56} & {19}
                  & {\tt 4$\cdot$9} & {12}                   \\
\hline
${\bar \nu}_{\mu} \nu_{\mu}$ & {\bf 10} & {\bf 10} & {20}
             & {\it 19} & {12} & {\tt 4$\cdot$3}  \\
\hline
${\bar \nu}_{\tau} \nu_{\tau}$ & {\bf 10} & {\bf 10} & {20}
             & {\bf 10} & {12} & {\bf 6}  \\
\hline
\end{tabular}
\vspace{.2cm} \\
\caption[]
{
Number of Feynman diagrams contributing to the
`NC' type production of two fermions pairs.
}
\end{center}
\vspace{-.5cm}
\end{table}
An example of the `NC' type process with 24 Feynman diagrams
(case~(i) of table 2) will be
discussed in another contribution to this conference~\cite{BLR94}.
For the more involved processes,
especially for case~(ii) of table 1
with final states containing an $e{\bar\nu_e}$ pair,
investigations have been started.
We expect
that the topologies indicated with roman numbers
may also be treated by our method.
\par
Another classification of four-fermion
processes was presented recently~\cite{BPK94}.
This reference separates leptonic, semileptonic,
and hadronic four-fermion final states.
For leptonic processes they agree with our number of Feynman diagrams.
For semileptonic processes agreement is found
in all cases but one, namely $q {\bar q} \nu_e {\bar \nu_e}$.
The authors of \cite{BPK94}
have 19 diagrams instead of our 21 for this case.
The explanation is
that for this process a $W$-fusion diagram exists, where
two $W$-bosons produce a $q {\bar q}$ state with a virtual $q^\prime$
in the $t$-channel.
For a given $q$, say $d$, $q^\prime$ may be $u,c$ or $t$
(due to Cabibbo-Kobayashi-Maskawa mixing).
In our classification we count all three
diagrams, while in \cite{BPK94} this is counted only once.
Finally, for hadronic processes we agree only in one of seven cases,
namely in our example with 11 diagrams.
The reasons for these differences are twofold.
Firstly, once more, quark mixing is neglected in \cite{BPK94}.
Secondly, we count gluon-exchange diagrams while this is not done
in~\cite{BPK94}.
Taking these differences into account, both
classifications agree.
\par
At the end of this section we introduce some notations.
Background diagrams may contain one resonating virtual $s$-channel
state or none (non-resonating background).
We will denote by a sub-index $n$
the total number of resonating $s$-channel propagators in a separate
contribution $\sigma_n$ to the cross-section of four-fermion
production,
\ba
e^+ e^-      &\rightarrow& (W^+ W^-,ZZ,ZH,\ldots) \rightarrow  4 f.
\nonumber
\label{eq4f}
\ea
Therefore, the basic contributions carry sub-index 4, the
background-basic interferences 3~or~2, and the pure
background contributions have sub-index 2,~1~or~0.
With appropriate kinematical cuts,
certain resonating states may be selected and
contributions with indices 4,~3,~2,~1, and 0 should be
hierarchically less and less important.
\section{BASIC CROSS-SECTIONS}
Here we present the basic cross-section formulae for the three cases
described by the diagrams of figure~1 with $WW$, $ZZ$, and
$ZH$ intermediate states.
They all may be expressed by twofold convolutions of a hard-scattering
off-shell cross-section with Breit-Wigner density functions.
For the $WW$ case, this representation
was invented in~\cite{MNW86}:
\ba
\sigma^{WW}(s) & = &
\int\limits_0^s ds_1 \, \rho_W(s_1)
\int\limits_0^{(\sqrt{s} - \sqrt{s_1})^2} ds_2 \, \rho_W(s_2)
\nl
& & \times~ \sigma^{WW}_4(s;s_1,s_2),
\label{sigww}
\ea
where
\ba
\rho_W(s_i)
=
\frac{1}{\pi}
\frac {\sqrt{s_i} \, \Gamma_W (s_i)
\times {\mr {BR}}(i)}
      {|s_i - M_W^2 + i \sqrt{s_i} \, \Gamma_W (s_i) |^2}
\label{rhow}
\ea
is the Breit-Wigner density function originating from the $W^{\pm}$
$s$-channel propagators
and BR($i$) is the
corresponding branching ratio.
Similar densities $\rho_Z$ and $\rho_H$ associated with $Z$ ($H$)
$s$-channel exchanges can be obtained by the replacements
$M_W, \Gamma_W \rightarrow M_Z, \Gamma_Z; (M_H, \Gamma_H $).
They are normalized so that
\bq
  \rho_B(s)
\stackrel {\Gamma_B \rightarrow 0} {\longrightarrow}
\delta(s - M_B) \times {\mr {BR}}(i), 
B=W,Z,H.
\label{normal}
\eq
Further,
\bq
\ww (s)
=
\frac{G_{\mu}\, M_W^2} {6 \pi \sqrt{2}}
\sqrt{s} \sum_f~N_c(f)
\label{gwoff}
\eq
is the off-shell ($s$-dependent) $W$-width and
$N_c(f)$=1(3) for leptons(quarks).
The sum in~(\ref{gwoff})
extends over all open fermion channels. At extremely
high energies, additionally opening channels may substantially
contribute to $\Gamma_W(s)$.
\par
The cross-section $\sigma^{WW}_4(s,s_1,s_2)$ contains six pieces but
is described by only three functions ${\cal G}_4^a(s;s_1,s_2)$,
which are different for the $s$- and $t$-channel and the
$st$-interference:
\ba
\lefteqn{\sigma^{WW}_4(s;s_1,s_2) =}
\nl
  & {\displaystyle \frac{\left(G_{\mu} M_W^2 \right)^2} {8 \pi s}}
  & \Biggl[ \Bigl( c_{\gamma \gamma} + c_{\gamma Z}
  + c_{ZZ} \Bigr)
  {\cal G}^{\mr s}_4(s;s_1,s_2)
\nl
  & & +~ \left(c_{\nu \gamma} + c_{\nu Z} \right)
      {\cal G}^{\mr {st}}_4(s;s_1,s_2)
\nl
  & & +~ c_{\nu \nu}  {\cal G}_4^{\mr t}(s;s_1,s_2) \Biggr].
\label{sigww4}
\ea
The coefficients $c_{\alpha\beta}$ consist of $Z(\gamma)ee$-
and $Z(\gamma)WW$-couplings and $\gamma,Z$-propagator ratios:
\ba
c_{\gamma \gamma} &=& 8 s_W^4 \, Q_e^2,
\nl
c_{\gamma Z} &=&
 4 s_W^2 \, v_e \, |Q_e| \, \Re e
\frac{s}{s-M_Z^2+i M_Z \Gamma_Z(s)},
\vspace*{3mm}
\nl
c_{ZZ} &=&
\frac{1}{2} \left( v_e^2 + a_e^2 \right)
\left| \frac{s}{s-M_Z^2+i M_Z \Gamma_Z(s)} \right|^2,
\nl
c_{\nu \gamma} &=& -4s_W^2 \, |Q_e|,
\nl
c_{\nu Z} &=& -(v_e+a_e) \, \Re e
\frac{s}{s-M_Z^2+i M_Z \Gamma_Z(s)},
\vspace*{3mm}
\nl
c_{\nu \nu} &=& 1,
\nonumber
\nl
\ea
with $Q_e=-1, a_e=1, v_e=1-4s_W^2, s_W^2=\sin^2 \theta_W$.
The three irreducible kinematical functions are
\ba
{\cal G}_4^{\mr s}(s;s_1,s_2) =
\frac{\lambda^{3/2}}{s^3 s_1 s_2} \left[ \frac{\lambda}{6} + 2s(s_1+s_2)
+2s_1s_2\right],
\nonumber
\ea
\ba
{\cal G}_4^{\mr {st}}(s;s_1,s_2)
=
\frac{\lambda^{1/2}}{s^2 s_1 s_2} \Biggl\{
\frac{\lambda}{6} \left[ s+11(s_1+s_2)\right]
\nl
+~2s(s_1^2 + 3s_1s_2 +s_2^2)
-2(s_1^3+s_2^3)
\nl
-~4s_1s_2 \left[ s(s_1+s_2)+s_1s_2\right] {\cal L}_4 \Biggr\},
\nonumber
\ea
\ba
{\cal G}_4^{\mr t}(s;s_1,s_2)
=
\frac{\lambda^{1/2}}{s s_1 s_2}
\Biggl[ \frac{\lambda}{6} + 2s(s_1+s_2)
\nl
-~8s_1s_2
+4s_1s_2 (s-s_1-s_2) {\cal L}_4\Biggr],
\label{wwmuta}
\ea
with
\ba
\lambda &\equiv&
 \lambda(s;s_1,s_2)
\nl
&=&
s^2 + s_1^2 + s_2^2 - 2ss_1 - 2 s_1s_2 - 2 s_2s
\label{lambda}
\ea
and
\ba
{\cal L}_4(s;s_1,s_2) &=&
\frac{1}{\sqrt{\lambda}} \, \ln \frac{s-s_1-s_2+\sqrt{\lambda}}
                                     {s-s_1-s_2-\sqrt{\lambda}}~.
\label{L4}
\ea
In the limit~(\ref{normal}),
the on-shell $W$-pair production cross-section
is obtained as given in~\cite{ONWW}.
\par
The corresponding set of formulae for basic off-shell $ZZ$ production is:
\ba
\sigma^{ZZ}(s) & = &
\int\limits_0^s ds_1 \, \rho_Z(s_1)
\int\limits_0^{(\sqrt{s} - \sqrt{s_1})^2} ds_2 \, \rho_Z(s_2)
\nl
& & \times~ \sigma^{ZZ}_4(s;s_1,s_2).
\label{sigzz}
\ea
The off-shell $Z$-boson width, which enters the definition of $\rho_Z$,
has the form
\bq
\Gamma_Z (s) =
  \frac{G_{\mu}\, M_Z^2} {24\pi \sqrt{2}} \sqrt{s}
  \sum_f (v_f^2+a_f^2) N_c(f).
\label{gzoff}
\eq
The cross-section $\sigma^{ZZ}_4(s,s_1,s_2)$ is extremely compact and
can be described by only one function ${\cal G}_4^{t+u}(s;s_1,s_2)$,
which is the sum of three others:
\ba
   {\cal G}_4^{\mr{t+u}}(s;s_1,s_2) &=& {\cal G}_4^{\mr t}(s;s_1,s_2)
  +~{\cal G}_4^{\mr{u}}(s;s_1,s_2)
\nl
&&+~{\cal G}_4^{\mr{tu}}(s;s_1,s_2).
\ea
Here the functions ${\cal G}_4^{\mr u}(s;s_1,s_2)
= {\cal G}_4^{\mr t}(s;s_2,s_1)$ correspond to $u$- and $t$-channel
diagrams and ${\cal G}_4^{tu}(s,s_1,s_2)$ describes the
$tu$-interference. The cross-section $\sigma^{ZZ}_4$ is given by:
\ba
\lefteqn{\sigma^{ZZ}_4(s;s_1,s_2) =}
\nl
\hspace*{-.5cm}
& {\displaystyle \frac{\left(G_{\mu} M_Z^2 \right)^2}{64\pi s}} \!
\left(v_e^4+6v_e^2a_e^2+a_e^4\right) {\cal G}_4^{\mr{t+u}}(s,s_1,s_2)
\label{sigzz4}
\ea
with the kinematical function
\ba
{\cal G}_4^{\mr{t+u}}(s;s_1,s_2) =
\frac{\lambda^{1/2}}{s}\left[\frac{s^2+(s_1+s_2)^2}{s-s_1-s_2}
  {\cal L}_4 -2 \right].
\nonumber 
\ea
Finally, the basic off-shell $ZH$~cross-section is:
\ba
\sigma^{ZH}(s) & = &
\int\limits_0^s ds_1 \, \rho_H(s_1)
\int\limits_0^{(\sqrt{s} - \sqrt{s_1})^2} ds_2 \, \rho_Z(s_2)
\nl
& & \times~ \sigma^{ZH}_4(s;s_1,s_2).
\label{sigzh}
\ea
Below the threshold of the decay $H \rightarrow W^+W^-$,
the off-shell width has the form
\bq
\Gamma_H (s) =
  \frac{G_{\mu}} {4\pi \sqrt{2}} \sqrt{s} \sum_f m_f^2 N_c(f).
\label{ghoff}
\eq
The cross-section $\sigma^{ZH}_4(s;s_1,s_2)$ is given by
\ba
\sigma^{ZH}_4(s;s_1,s_2)  =
\frac{\left(G_{\mu} M_Z^2 \right)^2} {96 \pi s}
\frac{M_Z^2}{s} \left( v_e^2+a_e^2 \right)
\nl
\times \left| \frac{s}{s-M_Z^2+i M_Z\Gamma_Z(s)} \right|^2
{\cal G}_4^{\mr{Bj}}(s;s_1,s_2).
\label{sigzh4}
\ea
Again, it contains only one kinematical
function, namely ${\cal G}_4^{\mr{Bj}}(s;s_1,s_2)$:

\bq
{\cal G}_4^{\mr {Bj}}(s;s_1,s_2) =
\frac{\lambda^{1/2}}{s^2 s_2}
\left(\lambda +12 s s_2\right).
\label{zhmuta}
\eq

The energy dependences of the off-shell $WW$, $ZZ$, and $ZH$
basic cross-sections are presented in figures~3--5.
For comparison we also show on-shell cross-sections
for all three cases. With respect to the on-shell case,
off-shell cross-sections are substantially
reduced in the threshold region and develop
tails at high energies so
that boson widths cannot be neglected for $s \gg (M_{B_1}+M_{B_2})^2$.
Going off-shell, the cross-section peaks are shifted to higher energies.
Using  constant widths, i.e. $\sqrt{s} \, \Gamma_B(s) \rightarrow M_B
\Gamma_B$, corresponds to redefinitions of the boson
masses~\cite{massdef}:
$M_B \rightarrow {\bar M}_B = M_B + \frac{1}{2} \Gamma_B^2/M_B$.
For the $W$ this results in ${\bar M}_W \approx M_W + 26$
MeV~\cite{BD94}.
That lowers the cross-section around threshold by at most $\approx -1.7\%$,
while for $\sqrt{s} > 180$ GeV the effect is small and positive
($\leq 0.1\% $). Numerical results were obtained with the Fortran
program {\tt GENTLE}~\cite{gentle}.
\begin{figure}[htbp]
\vspace{6cm}
\begin{center} \mbox{
\epsfysize=7.5cm
}
\end{center}
\caption[nonuniversal for ww.]
{The basic $W$-pair production.}
\label{fig3}
\end{figure}
%
\begin{figure}[hbtp]
\vspace{6cm}
\begin{center} \mbox{
\epsfysize=7.5cm
}
\end{center}
\caption[The basic off-shell $Z$-pair production.]
{The basic $Z$-pair production.}
\label{fig4}
\end{figure}
%
\begin{figure}[htbp]
\vspace{6cm}
\begin{center} \mbox{
\epsfysize=7.5cm
}
\end{center}
\caption[The basic off-shell $ZH$ production.]
{The basic $ZH$ production.}
\label{fig5}
\end{figure}
\section{BACKGROUND}
In this section we briefly sketch our results for
the semi-analytical treatment of the process~(\ref{eqabs}),
the production of four different fermions excluding electron
or electron neutrino (case~(i) in table 1).
This is described by the basic diagrams of figure~1 and by the eight
background diagrams of figure~2. In the unitary gauge, there are no
other diagrams for this particular final state.
All results were obtained with the help of {\tt FORM}~\cite{form},
{\tt SCHOONSCHIP}~\cite{schoonschip}, and {\tt CompHEP}~\cite{comphep}.
\par
After the analytical integration over five angular variables,
one arrives at a doubly-convoluted
representation for the cross-section:
\ba
\sigma^{4f}(s) =
\int\limits_0^s ds_1  \rho_W(s_1)
\int\limits_0^{(\sqrt{s} - \sqrt{s_1})^2} ds_2 \, \rho_W(s_2)
\nl
\!\!\!\times~[\sigma^{WW}_4(s;s_1,s_2) + \sigma^{4f}_3(s;s_1,s_2)
\nl
+ \sigma^{4f}_2(s;s_1,s_2) ].
\label{sig4f}
\ea

The basic contribution $\sigma^{WW}_4$ is given by (\ref{sigww4}).
The term $\sigma^{4f}_2$ corresponds to figure~2 and
$\sigma^{4f}_3$ are interference contributions.

We obtained explicit representations for the cross-sections
$\sigma^{4f}_3(s;s_1,s_2)$ and $\sigma^{4f}_2(s;s_1,s_2)$
in terms of seven new kinematical functions
(three functions ${\cal G}_3^a$ and four functions ${\cal G}_2^a$),
coupling constants and propagator ratios (similar to~(\ref{sigww4})).
The formulae for the cross-sections themselves are rather appealing,
while only three of the
kinematical functions are of the same compactness as those of the basic
processes.
Four interference functions
(two ${\cal G}_3^a$ and two ${\cal G}_2^a$)
could be written only with the aid of a cumbersome
polynomial presentation of the following type:
\ba
{\cal G}_{3}^{a} (s,s_1,s_2) =
\sqrt{\lambda} \sum_{i,j=0}^1 \left[s_1{\cal L}_3(s;s_1,s_2)\right]^i
\nl
\times~\left[s_1s_2{\cal L}_4(s;s_1,s_2)\right]^j {\cal P}_{ij}^{a}(s;s_1,s_2),
\label{g3def}
\ea
\ba
{\cal G}_{2}^{a} (s,s_1,s_2) =
\sqrt{\lambda} \sum_{i,j=0}^1 \left[s_1{\cal L}_3(s;s_1,s_2)\right]^i
\nl
\times~\left[s_2{\cal L}_3(s;s_2,s_1)\right]^j {\cal P}_{ij}^{a}(s;s_1,s_2),
\label{g2def}
\ea
\ba
{\cal P}_{ij}^{a}(s;s_1,s_2) &=&
p_0^{a}(ij)
\nl
&&+~\sum_{n=1}^3\frac{(s_1s_2)^{n-1}}{\lambda^n}
p_n^{a}(ij),
\label{g23def}
\ea
with ${\cal L}_4(s;s_1,s_2)$ given by~(\ref{L4}) and
\bq
{\cal L}_3(s;s_1,s_2) =
\frac{1}{\sqrt{\lambda}} \, \ln \frac{s+s_1-s_2+\sqrt{\lambda}}
                                     {s+s_1-s_2-\sqrt{\lambda}}.
\label{L3}
\eq
In~(\ref{g3def})
the $a$ stands for $U,D$,
and in~(\ref{g2def}) for $UD,U{\bar{U}}$.
The $p_n^{a}(ij), n=0,1,2,3$
in~(\ref{g23def}) are polynomials
of order $n$ in $s,s_1,s_2$.
A complete
analytical result will be presented elsewhere~\cite{isr2}.
Here we restrict ourselves to a comment and some numerical results.
As may be seen from~(\ref{g3def})--(\ref{g23def}),
the cumbersome interference functions
contain inverse powers of $\lambda$ (up to the third power), which
vanish at the upper limit of integration over $s_2$.
This is a typical example of so-called kinematical singularities.
Expanding ${\cal L}_{3,4}$ in Taylor series in $\lambda$, one may see
that all these
inverse powers cancel and the cross-section has a proper threshold
behaviour.
However, these kinematical singularities may create complications
for numerical calculations.

In figure~6, we present the ratios (basic+background)/(basic),
\bq
 R=\frac{\sigma^{4f}(s)} {\sigma^{WW}(s)},
\eq
for three different channels with 11, 10, and 9 diagrams respectively
(see case~(i) of table 1) as functions of ${\sqrt s}$.
As is seen from the figure,
the background contributions for these processes
are relatively small, especially at \ltwo\ energies.
Below threshold of the $W$-production and at high energies, the relative
contribution of such background increases and reaches $\sim 2 \%$.
\begin{figure}[thbp]
\vspace{6cm}
\begin{center} \mbox{
\epsfysize=7.5cm
}
\end{center}
\caption[background/basic for ww.]
{The (basic+background)/basic ratio $R$ for off-shell $W$-pair
  production with $l_i \neq e$.}
\label{fig6}
\vspace{.5cm}
\end{figure}
%
\section{INITIAL STATE RADIATIVE CORRECTIONS\break
{\kern-.2667em IN FOUR-FERMION PRODUCTION PROCESSES}}
Since the background contribution is comparatively small,
it is quite reasonable to restrict oneself
to the basic diagrams when calculating the complete lowest-order ISR
QED corrections.
\par
Let us recall that the ISR corrections are known to be
dominating in single $Z$-resonance production.
A similar property holds for
four-fermion processes at \ltwo,~which will proceed near
the corresponding $B_1B_2$-thresholds and the intermediate-state bosons
will be nearly at rest~\cite{wwqed1}.
\par
The ISR corrections for a cross-section
described by $s$-channel diagrams with $\gamma$-
and $Z$-exchanges (i.e. including the background terms) may be presented by
a universal formula~\cite{berends}-~\cite{BPKISR},
\ba
\sigma^{B_1B_2,{\mr s}}_{\mr{univ}}(s) =
\int\limits_0^s ds_1 \rho_{B_1}(s_1)
\int\limits_0^{(\sqrt{s} - \sqrt{s_1})^2} ds_2 \rho_{B_2}(s_2)
\nl
\int\limits_{(\sqrt{s_1}+\sqrt{s_2})^2}^s \frac{ds'}{s}
\left[ \beta_e v^{\beta_e - 1} (1+ \bar S) + {\bar H} \right]
\nl
\times~\sigma_{4}^{B_1B_2}(s';s_1,s_2),
\nonumber
\ea
\ba
\label{qed-s}
\ea
where $v=1-s'/s$.
The soft plus virtual photon part $\bar S$ and the hard part
${\bar H} (s'/s)$ are given by
\ba
{\bar S}&=&\frac{\alpha}{\pi}
\left[ \frac{\pi^2}{3} - \frac{1}{2} \right]
+ \frac{3}{4} \beta_e + {\cal O}(\alpha^2),
\\
{\bar H}(s'/s) &=& - \frac{1}{2} \left(1+\frac{s'}{s}\right)
\beta_e + {\cal O}(\alpha^2),
\label{betae}
\ea
and $\beta_e=2 \alpha/\pi [ \ln (s/m_e^2) - 1 ]$.
\par
Equation~(\ref{qed-s}) may be directly applied to the case $B_1B_2=ZH$
and to the $s$-channel contribution of $B_1B_2=WW$. We have rederived
by explicit calculations
that~(\ref{qed-s}) may be obtained straightforwardly from
the usual vertex and bremsstrahlung QED Feynman diagrams
after up to seven sequential angular integrations
(five for the vertex part and seven for bremsstrahlung).
The situation becomes much more complicated
if $t$- and $u$-channel exchange diagrams are involved.
Here we face two kinds of problems.
The factorized form~(\ref{qed-s})
is no longer valid for the squared $t$- and $u$-channel
diagrams and for $st$-,$su$-, and $tu$-interferences.
One of the reasons is the angular dependence of $t$- and $u$-channel
propagators.
This leads to the appearance of additional non-factorizable
(non-universal) QED corrections.
\par
For the $ZZ$ process this is the only problem, although
technical complications arise due to additional Feynman diagrams with
real and virtual photons attached
to the virtual electron line in $t$- and $u$-channels.
For the $WW$ basic diagrams another problem persists.
There, the electric
charge flows from the initial state electron through an intermediate
$W$-boson to a final state fermion.
Therefore, only the complete set of all QED diagrams is gauge-invariant.
A `\naive' subset of diagrams with ISR corrections
corresponding to the diagrams with a (real or virtual) photon
attached to the external electron legs is not
gauge-invariant.
This is different from the $ZZ$ basic diagrams, where the
electric charge flows continuously through the initial state.

A straightforward solution of the problem would be a complete numerical
calculation of {\em all} the \oalf\ corrections,
including also intermediate and
final state corrections.
This was done for the on-shell case in~\cite{jegerlehner},
but it seems to be incredibly complicated
for the off-shell case (see~\cite{aeppli}).
In such a complete approach one must also include all electroweak
corrections and properly treat
the radiation from virtual intermediate $W^{\pm}$-states.
\par
We used a completely different approach to the problem,
namely to take advantage of the
fact that ISR corrections should yield the main fraction of the
net correction.
Therefore, we tried to define a gauge-invariant ISR correction
by splitting the electrically neutral neutrino flow in the $t$-channel
into two oppositely flowing charges \mbox{--1} and \mbox{+1}.
The charge --1 is then combined with the `\naive' ISR diagrams
in order
to build a continuous flow of electric charge in the initial state.
The charge +1 is combined with the intermediate-state photon
emission and is neglected here.
This technique, which is explained in more
detail in~\cite{wwqed1}, is called
the {\em Current Splitting Technique} (CST).
\par
Within the CST, the $ZZ$ and $WW$ $t$-channel QED amplitudes
are identical.
The difference between the two cases arises at the
level of cross-section calculations.
For the $WW$ case, there are $st$ and $tt$ interferences, while
for the $ZZ$ case one finds $t$- and $u$-channel contributions and the
$tu$ interference. In all cases the cross-section has the following
structure:
\ba
\frac{d^3\sigma^{B_1B_2,{\mr a}}_{\mr{non-univ}}(s)}{ds_1ds_2ds'}
& = &
\frac{1}{s}\rho_{B_1}(s_1) \rho_{B_2}(s_2)
\nl
&&\times \left[ \beta_e v^{\beta_e - 1} {\cal S}_{\mr a}
+{\cal H}_{\mr a} \right],
\label{qed-t}
\ea
with
\ba
{\cal S}_{\mr a}(s,s';s_1,s_2)
&=&
\left[ 1 + {\bar S}(s) \right] \sigma_0^{B_1B_2,{\mr a}}(s';s_1,s_2)
\nl
&&
+~\sigma^{B_1B_2,{\mr a}}_{\hat S}(s';s_1,s_2),
\label{hats}
\ea
\ba
{\cal H}_{\mr a}(s,s';s_1,s_2)
&=&
{\bar H}(s,s') \sigma_0^{B_1B_2,{\mr a}}(s';s_1,s_2)
\nl
&&
+~\sigma^{B_1B_2,{\mr a}}_{\hat H}(s,s';s_1,s_2),
\label{hath}
\ea
where
$\sigma^{B_1B_2,{\mr a}}_{\hat S}(s';s_1,s_2)$ and
$\sigma^{B_1B_2,{\mr a}}_{\hat H}(s,s';s_1,s_2)$ are non-universal,
non-factorizable soft and hard contributions.
Equation~(\ref{qed-t}) possesses several remarkable properties.
The leading ISR correction contributions to the cross-section,
containing mass singularities via $\beta_e$,
factorize for any $a=st,su,tu$-interferences and
$t$- and $u$-channel exchanges. This is necessary to
ensure that the gauge cancellation is not spoiled.
\par
The non-universal terms are calculated so far only for the
$WW$-case~\cite{wwqed1}.
As one should expect, the non-universal contributions
do not contain mass singularities.
An analogous study for the $ZZ$ case is in progress~\cite{BLR}.
\par
The numerical influence of the universal part of the ISR corrections
(i.e. setting non-universal parts equal to zero) is presented in
figures~3--5 for the $WW$, $ZZ$, and $ZH$ cases.
Universal ISR yields large, negative contributions in the vicinity of
the threshold. At high energies  these corrections are positive
and one observes
the effect of the radiative tail similar to
the $Z$-peak. This radiative tail phenomenon is more pronounced than
the high energy tail due to the bosons' off-shellness. We do not show
the radiatively corrected on-shell cross-section
$\sigma_{\mr {univ}}^{\mr{on}}$, but only mention that the relative
differences between $\sigma_{\mr {Born}}^{\mr{on}}$ and
$\sigma_{\mr {Born}}^{\mr{off}}$ on the one hand and between
$\sigma_{\mr {univ}}^{\mr{on}}$ and $\sigma_{\mr {univ}}^{\mr{off}}$
on the other are similar.
\begin{figure}[htbp]
\vspace{6cm}
\begin{center} \mbox{
\epsfysize=7.45cm
}
\end{center}
\caption[nonuniversal ww.]
{The non-universal initial state and Coulomb
corrections to off-shell $W$-pair
production.}
\label{fig7}
\end{figure}
\par
The effect of the non-universal
contributions in the $WW$-case is illustrated
in figure~7 (solid line).
This contribution is seen to be small, at \ltwo\ energies
it does not exceed $0.4 \%$.
At high energies the relative
contribution of the non-universal term becomes as large as
1.4\% at $\sqrt{s}=1$ TeV.
To a great extent the smallness of the non-universal terms
at high energies
 is due to
the {\em screening} property of the non-universal corrections.
They have a damping overall factor,
\ba
\sigma^{\mr {st,t}}_{{\hat S},{\hat H}}(s';s_1,s_2)
\sim
\frac{s_1 s_2}{s^2}.
\label{scree}
\ea
The screening property ensures the unitary behaviour of the
non-universal terms at high energy for the individual ($st$ and $t$)
contributions.
\par
In figure~7, we also show an important part of the
final state corrections -- the so-called Coulomb singularity.
It yields a positive correction,
which has its maximum value of about 6\% at the threshold and
vanishes at high energies. At $\sqrt{s}$=1000~GeV it amounts to 0.75\%.
The Coulomb correction is taken into account
according to equation~(5) of~\cite{BBD}.
However, at high energies,
other final state corrections are important~\cite{jegerlehner}.
%
\section{CONCLUSIONS AND PROSPECTS}
For a variety of purposes, the semi-analytical treatment of four-fermion
production is an interesting alternative to the Monte Carlo approach.
\\
Within the {\tt GENTLE} project, we have calculated so far
\begin{itemize}
\item[(i)]
the \oa\ ISR corrections to and the average radiative energy loss
$\langle E_{\rm {rad}} \rangle$ in
 off-shell $W$-pair production;
\item[(ii)]
the background contribution for this process with the simplest final state
configuration;
\item[(iii)]
the off-shell $ZH$ production with universal ISR corrections and the background
contributions for the $\bar \mu \mu \bar b b$ decay mode.
\end{itemize}
We are presently studying
\begin{itemize}
\item[(iv)]
the \oa\ ISR corrections to off-shell $Z$-pair production.
\end{itemize}
We intend to study
\begin{itemize}
\item[(v)]
background contributions to $Z$- and $W$-pair production with other
topologies;
\item[(vi)]
background contributions with $t$-channel exchanges and final states with
$e^{\pm}$ and $\nu_e$ or $\bar \nu_e$.
\item[(vii)]
final state QED corrections to on-shell
$W$-pair production with the current-splitting technique;
\item[(viii)]
the inclusion of virtual weak corrections.
\end{itemize}

The annihilation of two particles into four
(five particles in the case of real\break bremsstrahlung corrections)
has a limited variety of topologies in the tree approximation for the basic
process.
Thus, its systematic treatment is of principal theoretical interest.
Apart from the case of \ltwo, it finds additional applications in e.g. the
study of the $Z$ line shape at
LEP~1 (initial state QED or final state QED and QCD higher order pair
production corrections) or in QCD corrections at the LHC.
A study of certain problems beyond the Standard Model is, of course, also
within reach.

\end{document}